\documentclass[twoside,11pt]{article}

\usepackage[final]{melba}

\usepackage{amsmath,amsfonts}
\usepackage{graphicx}
\usepackage[noend]{algpseudocode}
\usepackage{algorithm,algorithmicx}
\usepackage{amsmath}
\usepackage{amssymb}
\usepackage{rotating}
\usepackage{multirow}
\usepackage{todonotes}
\usepackage{graphicx}
\usepackage[export]{adjustbox}
\usepackage{xcolor}
\usepackage{caption}
\usepackage{bm}
\usepackage{threeparttable}
\usepackage{placeins}
\usepackage[normalem]{ulem}
\usepackage{float}
\usepackage{wrapfig}

\def\UL#1{\expandafter\nusp#1 \nil}
\def\nusp#1 #2\nil{\sout{#1} \ifx\relax#2\relax\else\nusp#2\nil\fi}

\newcommand{\bb}[1]{\bm{\mathrm{#1}}}

\melbaheading{6}{https://www.melba-journal.org/article/21965}{2021}{1-23}{12/2020}{3/2021}{Weiss, Senouf, Vedula, Michailovich, Zibulevsky and Bronstein}{}{}

\ShortHeadings{PILOT: Physics-Informed Learned Optimized Trajectories for Accelerated MRI}{Weiss et al.}
\firstpageno{1}

\title{PILOT: Physics-Informed Learned Optimized Trajectories for Accelerated MRI}

\author{\name Tomer Weiss \email tomer-weiss@cs.technion.ac.il \\  
	\addr Computer Science, Technion, Haifa, Israel
	\AND
	\name Ortal Senouf \email senouf@cs.technion.ac.il \\
	\addr Computer Science, Technion, Haifa, Israel
	\AND
	\name Sanketh Vedula \email sanketh@cs.technion.ac.il \\
	\addr Computer Science, Technion, Haifa, Israel
	\AND
	\name Oleg Michailovich \email olegm@uwaterloo.ca \\
	\addr Electrical and Computer Engineering, Waterloo University, Waterloo, Canada
	\AND
	\name Michael Zibulevsky \email mzib@cs.technion.ac.il \\
	\addr Computer Science, Technion, Haifa, Israel
	\AND
	\name Alex Bronstein \email bron@cs.technion.ac.il \\
	\addr Computer Science, Technion, Haifa, Israel
}

\begin{document}

\maketitle

\begin{abstract}
Magnetic Resonance Imaging (MRI) has long been considered to be among ``the gold standards" of diagnostic medical imaging. The long acquisition times, however, render MRI prone to motion artifacts, let alone their adverse contribution to the relative high costs of MRI examination. Over the last few decades, multiple studies have focused on the development of both physical and post-processing methods for accelerated acquisition of MRI scans. These two approaches, however, have so far been addressed separately. On the other hand, recent works in optical computational imaging have demonstrated growing success of {\it concurrent learning-based design} of data acquisition and image reconstruction schemes. Such schemes have already demonstrated substantial effectiveness, leading to considerably shorter acquisition times and improved quality of image reconstruction. Inspired by this initial success, in this work, we propose a novel approach to the learning of optimal schemes for conjoint acquisition and reconstruction of MRI scans, with the optimization carried out simultaneously with respect to the time-efficiency of data acquisition and the quality of resulting reconstructions. To be of a practical value, the schemes are encoded in the form of general $k$-space trajectories, whose associated magnetic gradients are constrained to obey a set of predefined hardware requirements (as defined in terms of, e.g., peak currents and maximum slew rates of magnetic gradients). With this proviso in mind, we propose a novel algorithm for the end-to-end training of a combined acquisition-reconstruction pipeline using a deep neural network with differentiable forward- and back-propagation operators. We validate the framework restrospectively and demonstrate its effectiveness on image reconstruction and image segmentation tasks, reporting substantial improvements in terms of acceleration factors as well as the quality of these tasks. 

Code for reproducing the experiments is available at \url{https://github.com/tomer196/PILOT}.
\end{abstract}


\section{Introduction}

Magnetic Resonance Imaging (MRI) is a principal modality of modern medical imaging, particularly favoured due to its noninvasive nature, the lack of harmful radiation, and superb imaging contrast and resolution. However, its relatively long image acquisition times adversely affect the use of MRI in multiple clinical applications, including emergency radiology and dynamic imaging. 

In \cite{lustig2007sparse}, it was demonstrated that it is theoretically possible to accelerate MRI acquisition by randomly sampling the $k$-space (the frequency domain where MR images are acquired). However, many compressed sensing (CS) based approaches have some practical challenges; it is difficult to construct a feasible trajectory from a given random sampling density, or choose the $k$-space frequencies under the MRI hardware constraints.
In addition, the reconstruction of a high-resolution MR image from undersampled measurements is an ill-posed inverse problem where the goal is to estimate the latent image $\bb{Z}$ (fully-sampled $k$-space volume) from the observed measurements $\bb{X} = \mathcal{F}(\bb{Z})+\bb{\eta}$, where $\mathcal{F}$ is the forward operator (MRI acquisition protocol) and $\bb{\eta}$ is the measurement noise. Some previous studies approached this inverse problem by assuming \textit{priors} (typically, incorporated in a \textit{maximum a posteriori} setting) on the latent image such as low total variation or sparse representation in a redundant dictionary \citep{lustig2007sparse}. Recently, deep supervised learning-based approaches have been in the forefront of MRI reconstruction \citep{hammernik2018learning,sun2016deep,zbontar2018fastmri, multiMRI-GAN, dar2020transfer}, solving the above inverse problem through implicitly learning the prior from a data set, and exhibiting significant improvement in the image quality over the explicit prior methods.
Other methods, such as SPARKLING \citep{sparkling2019Lazarus}, have attempted to optimize directly the acquisition protocol, i.e, over the feasible $k$-space trajectories, showing further sizable improvements. The idea of joint optimization of the forward (acquisition) and the inverse (reconstruction) processes has been gaining interest in the MRI community for learning sampling patterns \citep{bahadir2019learning} and Cartesian trajectories \citep{our2019fastmri,gozcu2018learning,zhang2019reducing,aggarwal2020jmodl}.

\subsection{Related Works}
\cite{lustig2007sparse} showed that variable-density sampling pattern, with a gradual reduction in the sampling density towards the higher frequencies of the $k$-space, performs greatly when compared to uniform random sampling density suggested by the theory of compressed sensing \citep{candes2006stable}. The sampling pattern can be optimised in a given setting to improve reconstruction quality. We distinguish the recent works that perform optimization of sampling patterns into four paradigms:
\begin{enumerate}
    \vspace{-0.1cm}
    \setlength\itemsep{-0.15cm}
    \item designing 2D Cartesian trajectories, i.e., designing sampling patterns along the phase-encoding direction while fully sampling the frequency-encoding direction \citep{our2019fastmri,gozcu2018learning,zhang2019reducing, aggarwal2020jmodl, sanchez2020dynamic}; 
    \item designing 2D variable-density sampling patterns and performing a full Cartesian sampling along the third dimension \citep{bahadir2019mri};
    \item designing {feasible} non-Cartesian 2D trajectories \citep{sparkling2019Lazarus}; 
    \item designing {feasible} non-Cartesian 3D trajectories \citep{lazarus20193d, alush20203dflat}.
\end{enumerate}

Our work falls into the third paradigm, and it exploits the complete degrees of freedom available in 2D. 
Parallel line of works focused on finding the best sampling density to sample from \citep{lustig2010spirit, senel2019sampling, knoll2011adapted}. On the theoretical front, the compressed sensing assumptions have been refined to derive optimal densities for variable density sampling \citep{puy2011variable, chauffert2013variable, chauffert2014tsp}, to prove bounds on reconstruction errors for variable density sampling \citep{adcock2017breaking, krahmer2013stable} and to prove exact recovery results for Cartesian line sampling \citep{boyer2019compressed}. As \cite{boyer2019compressed} points out, approximating non-uniform densities with 2D Cartesian trajectories is demanding and requires high dimensions. 
On the other hand, acquiring non-Cartesian trajectories in $k$-space is challenging due to the need of adhering to physical constraints imposed by the machine, namely the maximum slew rate of magnetic gradients and the upper bounds on the peak currents. \\

Addressing the above problem has yielded a number of interesting solutions, which have effectively extended the space of admissible trajectories to include a larger class of parametric curves, with spiral \citep{spiral1992, singleshotspiral2005}, radial \citep{radial1973} and rosette \citep{rosette1981} geometries being among the most well-known examples. Some of these challenges have been recently addressed by the SPARKLING trajectories proposed in \cite{sparkling2019Lazarus}. However, this solution does not exploit the strengths of data-driven learning methods, merely advocating the importance of constrained optimization in application to finding the trajectories that best fit \textit{predefined} sampling distributions and contrast weighting constraints, subject to some additional hardware-related constraints.
The main drawback of this approach is, therefore, the need to heuristically design the sampling density.
 
To overcome the above limitations, in our present work, we suggest a new method of cooperative learning which is driven by the data acquisition, reconstruction, and end task-related criteria. In this case, the resulting $k$-space trajectories and the associated sampling density are optimized for a particular end application, with image reconstruction and image segmentation being two important standard cases that are further exemplified in our numerical experiments. 

\subsection{Contribution}
This paper makes three main contributions: 
Firstly, we introduce PILOT (\textbf{P}hysics \textbf{I}nformed \textbf{L}earned \textbf{O}ptimized \textbf{T}rajectory), a deep-learning based method for joint optimization of physically-viable $k$-space trajectories. To the best of our knowledge, this is the first time when the hardware acquisition parameters and constraints are introduced into the learning pipeline to model the $k$-space trajectories conjointly with the optimization of the image reconstruction network. Furthermore, we demonstrate that PILOT is capable of producing significant improvements in terms of time-efficiency and reconstruction quality, while guaranteeing the resulting trajectories to be physically feasible.

Secondly, when initialized with a parametric trajectory, PILOT has been observed to produce trajectories that are (globally) relatively close to the initialization. A similar phenomenon has been previously observed in \cite{sparkling2019Lazarus, ours2019learningBF}. As a step towards finding globally optimal trajectories for the single-shot scenario, we propose two greedy algorithms: (i) Multi-scale PILOT that performs a coarse-to-fine refinement of the $k$-space trajectory while still enforcing machine constraints; (ii) PILOT-TSP, a training strategy that uses an approximated solution to the traveling salesman problem (TSP) to globally update the $k$-space trajectories during learning. We then demonstrate that multi-scale PILOT and PILOT-TSP provide additional improvements when compared to both basic PILOT and parametric trajectories.

Lastly, we further demonstrate the effectiveness of PILOT algorithms through their application to a different end-task of image segmentation (resulting in task-optimal trajectories). Again, to the best of our knowledge, this is the first attempt to design $k$-space trajectories in a physics-aware task-driven framework.

To evaluate the proposed algorithms, we perform extensive experiments through \textit{retrospective} acquisitions on a large-scale multi-channel raw knee MRI dataset \citep{zbontar2018fastmri} for the reconstruction task, and on the brain tumor segmentation dataset \citep{MRIsegment} for the segmentation task. We further study the impact of SNR of the received RF signal on the end-task, and demonstrate how \textit{noise-aware} trajectory learning robustifies the learned trajectories.

The remainder of this paper is organized as follows. In Section \ref{sec:Methods}, the proposed methods PILOT and PILOT-TSP are presented. Section \ref{sec:experiments} summarises the experiments conducted and discusses the results. Finally, the main conclusions are presented in Section \ref{sec:Conclusion}.

\section{Methods}
\label{sec:Methods}

\begin{figure*}[t]
	\centering
	\includegraphics[width=1\textwidth]{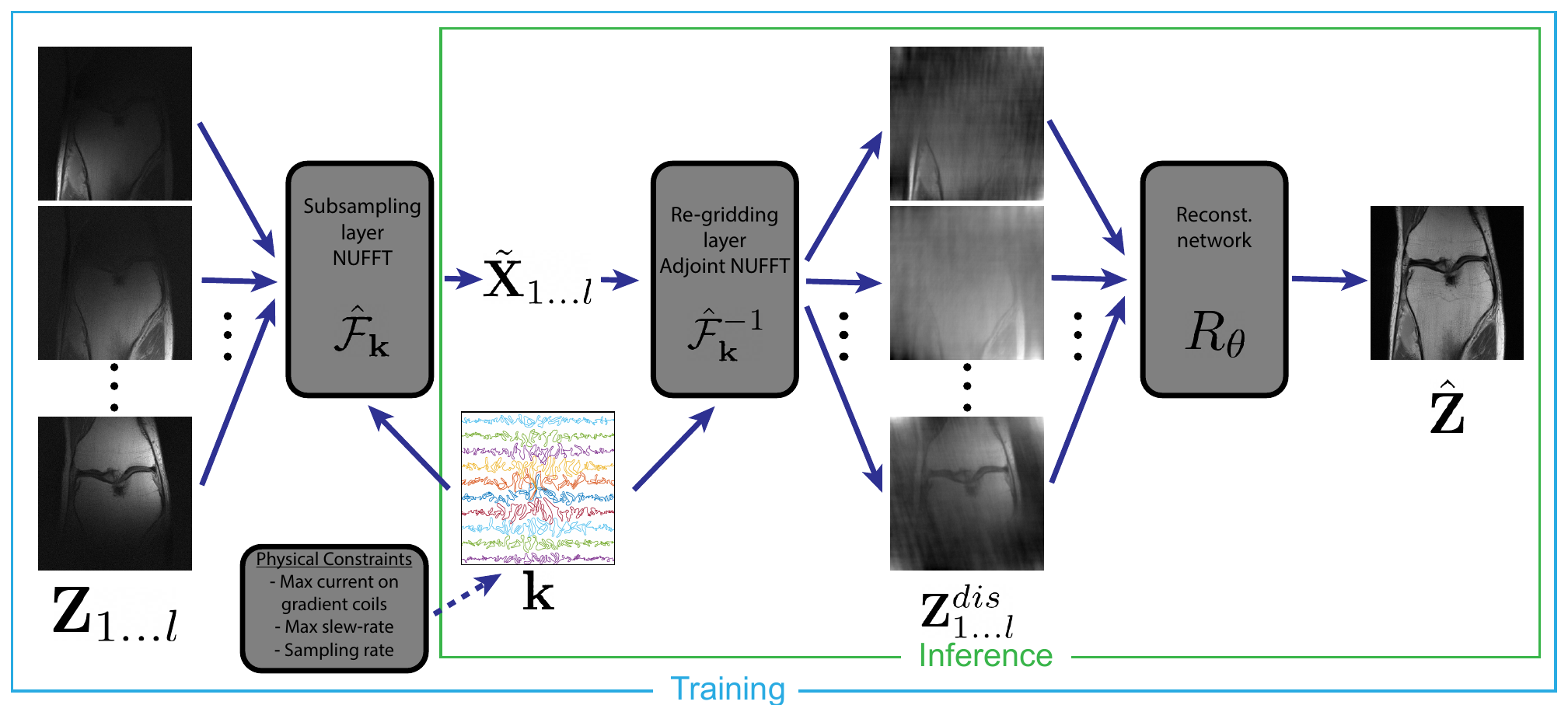} 
	\vspace{-0.3cm}
	\caption{\small{\textbf{PILOT pipeline.} Fully sampled multicoil images are inputs to the sub-sampling layer (NU-FFT) which samples the $k$-space along a trajectory $\mathbf{k}$. The measurements at the selected discrete set of frequencies are then passed to the regridding layer which performs a non-uniform Fast Fourier transform producing a space-domain image on a Cartesian grid. This image is then sent to the inverse model implemented as a reconstruction network (parameterized by $\bb{\theta}$). Physical constraints on the trajectory $\bb{k}$ are enforced at training.}}\label{fig:pipeline}
\vspace{-0.5cm}
\end{figure*} 

It is convenient to view our approach as a single neural network combining the forward (acquisition) and the inverse (reconstruction) models (see Fig. \ref{fig:pipeline} for a schematic depiction). The inputs to the model are the complex multi-channel fully sampled images, denoted as $\bb{Z}$ with $l$ channels. The input is faced by a sub-sampling layer modeling the data acquisition along a $k$-space trajectory, a regridding layer producing an image on a Cartesian grid in the space domain, and an end-task model operating in the space domain and producing a reconstructed image (reconstruction task) or a segmentation mask (segmentation task). 

In what follows, we detail each of the three ingredients of the PILOT pipeline. It should be mentioned that all of its components are differentiable with respect to the trajectory coordinates, denoted as $\bb{k}$, in order to allow training the latter with respect to the performance of the end-task of interest.
\subsection{Sub-sampling layer}
\label{subsec:subsampling}
The goal of the sub-sampling layer, denoted as $\hat{\mathcal{F}}_{\bb{k}}$, is to create the set of measurements to be acquired by each one of the MRI scanner channels along a given trajectory in $k$-space. The trajectory is parametrized as a matrix $\bb{k}$ of size $N_\textrm{shots}\times m$, where $N_\textrm{shots}$ is the number of RF excitations and $m$ is the number of sampling points in each excitation. The measurements themselves form a complex matrix of the same size emulated by means of non-uniform FFT (NU-FFT) \citep{Dutt1993NUFFT}
$\tilde{\bb{x}}_i = \hat{\mathcal{F}}_{\bb{k}} (\bb{Z}_i)$ on the full sampled complex images $\bb{Z}_i \in \mathbb{C}^{n \times n}$ for each one of the $l$ channels ($1\leq i\leq l$).
In the single-shot scenario, we refer to the ratio $n^2/m$ as the \emph{decimation rate} (\textrm{DR}).

\subsection{Regridding layer}
For transforming non-Cartesian sampled MRI $k$-space measurements to the image domain we chose to use the adjoint non-uniform FFT \citep{Dutt1993NUFFT}, henceforth denoted as $\hat{\mathcal{F}}^{*}_{\bb{k}}$. The non-uniform Fourier transform first performs regridding (resampling and interpolation) of the irregularly sampled points onto a regular grid followed by standard FFT. The result is a (distorted) MR image,
$\bb{Z}_i^\textrm{dis} = \hat{\mathcal{F}}^{*}_{\bb{k}} (\tilde{\bb{x}}_i)$ for $1\leq i\leq l$. 

Both the sub-sampling and the regridding layer contain an interpolation step. While it is true that the sampling operation is not differentiable (as it involves rounding non-integer values), meaningful gradients propagate through the bilinear interpolation module (a weighted average operation). For further details we refer the reader to Appendix A.

\subsection{Task model}
The goal of the task model is to extract the representation of the input images $\bb{Z}_{1...l}^\textrm{dis}$ that will contribute the most to the performance of the end-task such as reconstruction or segmentation. At training, the task-specific performance is quantified by a loss function, which is described in Section \ref{subsec:loss}.  
The model is henceforth denoted as $\hat{\bb{Z}} = R_{\bb{\theta}}(\bb{Z}_{1...l}^\textrm{dis})$, with $\bb{\theta}$ representing its learnable parameters. The input to the network is the distorted MR images, $\bb{Z}_{1...l}^\textrm{dis}$, while the output varies according to the specific task. For example, in reconstruction, the output is an MR image (typically, on the same grid as the input), while in segmentation it is a mask representing the segments of the observed anatomy.

In the present work, to implement the task models we used a root-sum-of-squares layer \citep{larsson2003sos} followed by a multi-resolution encoder-decoder network with symmetric skip connections, also known as the U-Net architecture \citep{ronneberger2015u}. U-Net has been widely-used in medical imaging tasks in general as well as in MRI reconstruction \citep{zbontar2018fastmri} and segmentation \citep{MRIsegment} in particular\footnote{We tried more advanced reconstruction architectures involving complex convolutions \citep{WANG2020136} that work on the individual complex channels separately but in our experiments complex convolutional networks yielded inferior performance comparing to U-Net working on single real channel post aggregating the multi-channel data.}. It is important to emphasize that the principal focus of this work is not on the task model \emph{in se}, since the proposed algorithm can be used with any differentiable model to improve the end task performance.

\subsection{Physical constraints}
A feasible sampling trajectory must follow the physical hardware constraint of the MRI machine, specifically the peak-current (translated into the maximum value of imaging gradients $G_\mathrm{max}$), along with the maximum slew-rate $S_{max}$ produced by the gradient coils. These requirements can be translated into geometric constraints on the first- and second-order derivatives of each of the spatial coordinates of the trajectory,
$
| \dot{k} | \approx \frac{| k_{i+1} - k_i |}{dt}  \le v_\mathrm{max} = \gamma\, G_\mathrm{max},
$
and 
$
| \ddot{k} | \approx \frac{| k_{i+1} - 2k_i + k_{i-1} | }{dt^2}  \le a_\mathrm{max} = \gamma \, S_\mathrm{max} ,
$
where $\gamma$ is the appropriate gyromagnetic ratio.

\subsection{Loss function and training}
\label{subsec:loss}

The training of the proposed pipeline is performed by simultaneously learning the trajectory $\bb{k}$ and the parameters of the task model $\bb{\theta}$. A loss function is used to quantify how well a certain choice of the latter parameters performs, and is composed of two terms: a task fitting term and a constraints fitting term,
$L = L_\text{task} + L_\text{const}.$

The aim of the first term is to measure how well the specific end task is performed. In the supervised training scenario, this term penalizes the discrepancy between the task-depended output of the model, $\hat{\bb{Z}}$, and the desired ground truth outcome $\bb{Z}$. For the reconstruction task, we chose the $L_1$ norm to measure the discrepancy between model output image $\hat{\bb{Z}}$ and the ground-truth image $\bb{Z}$, derived using the Root-Sum-of-Squares reconstruction \citep{larsson2003sos} of the fully sampled multi-channel images,
$L_\text{task} = \| \hat{\bb{Z}} - \bb{Z} \|_1$.
Similarly, for the segmentation task, we chose to use the cross-entropy operator to measure the discrepancy between the model output and the ground truth. This time, the model output $\hat{\bb{Z}}$ was the estimated segmentation map while $\bb{Z}$ is the ground truth segmentation map, The task loss is defined to be $ L_\text{task} = H(\hat{\bb{Z}}, \bb{Z})$, where $H$ is the cross-entropy loss.
As was the case for the task model, our goal was not to find an optimal loss function for the task-fitting term, having in mind that the proposed algorithm can be used with more complicated loss function such as SSIM loss \citep{zhao2016ssim-loss} or a discriminator loss \citep{GAN-loss@2018}, as long as the latter is differentiable.
 The second term $L_\text{const}$ in the loss function applies to the trajectory $\bb{k}$ only and penalizes for the violation of the physical constraints. We chose the hinge functions of the form $\mathrm{max}(0, |\dot{k}|-v_\mathrm{max})$ and $\mathrm{max}(0, |\ddot{k}|-a_\mathrm{max})$ summed over the trajectory spatial coordinates and over all sample points. These penalties remain zero as long as the solution is feasible and grow linearly with the violation of each of the constraints. The relative importance of the velocity (peak current) and acceleration (slew rate) penalties is governed by the parameters $\lambda_v$ and $\lambda_a$, respectively.

The training is performed by solving the optimization problem 
\begin{equation}
\min_{\bb{k}, \bb{\theta}} \, \sum_{(\bb{Z_{1...l}},\bb{Z})} L_\text{task}( R_{\bb{\theta}}(\hat{\mathcal{F}}^{*}_{\bb{k}}(\hat{\mathcal{F}}_{\bb{k}} (\bb{Z_{1...l}}) ) ) , \bb{Z}) + L_\text{const}(\bb{k}),
\label{eq:min}
\end{equation}
where the loss is summed over a training set comprising the pairs of fully sampled data $\bb{Z_{1...l}}$ and the corresponding groundtruth output $\bb{Z}$.
%

\subsection{Extensions}
Since optimization problem (\ref{eq:min}) is highly non-convex, iterative solvers are guaranteed to converge only locally. Indeed, we noticed that the learned trajectory $\bb{k}$ significantly depends on the initialization. To overcome this difficulty we propose the following two methods to improve the algorithm exploration and finding solutions closer to the global minimum of Eq. {\ref{eq:min}}.

\begin{figure*}[t]
	\centering
	\includegraphics[width=\textwidth]{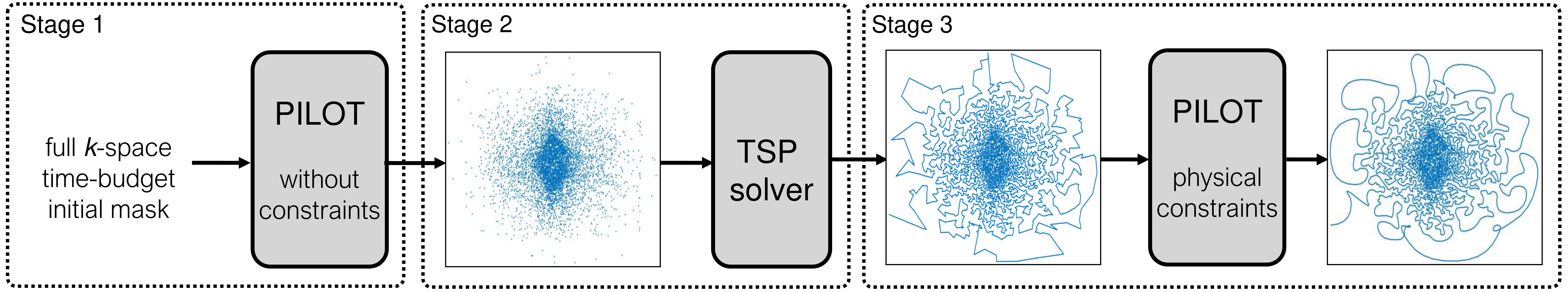} 
	\caption{\small{\textbf{PILOT-TSP training pipeline.} At the first stage, an unordered set of $k$-space points is learned without enforcing any hardware constraints. At the second stage, a greedy solver to the traveling salesman problem is employed to construct a trajectory passing through the sampled points. At the final stage, the obtained trajectory is used as the initialization the PILOT training pipeline with physical constraints enforced during training.}}\label{fig:TSP-pipeline}
\end{figure*}

\subsubsection{PILOT-TSP}
\label{sec:pilot-tsp}
A naive way to mitigate the effect of such local convergence is to randomly initialize the trajectory coordinates. However, our experiments show that random initialization invariably result in sequences of points that are far from each other and, thus, induce huge velocities and accelerations violating the constraints by a large margin. The minimization of the previously described loss function with such an initialization appears to be unstable and does not produce useful solutions. 

To overcome this difficulty, we introduce PILOT-TSP, an extension of PILOT for optimizing single-shot trajectories from random initialization. TSP stands for the Traveling Salesmen Problem, in our terminology, TSP aims at finding an ordering of $m$ given $k$-space coordinates such that the path connecting them has minimal length. Since TSP is NP-hard, we used a greedy approximation algorithm \citep{TSP1985} for solving it, as implemented in \texttt{TSP-Solver}\footnote{https://github.com/dmishin/tsp-solver}. Using this approach, we first optimize for the best unconstrained solution (i.e., the weights of the $L_\mathrm{const}$ term in the loss are set to zero), and then find the closest solution satisfying the constraints. 

The flow of PILOT-TSP, described schematically in Fig. \ref{fig:TSP-pipeline}, proceeds in four consecutive stages, as follows:
\begin{enumerate}

  \item Trajectory coordinates $\bb{k}$ are initialized by sampling $m$ i.i.d. vectors from a normal distribution with zero mean and standard deviation of $\sigma = \frac{n}{6}$; 
  
  \item The parameters $\bb{k}$ of the trajectory and $\bb{\theta}$ of the task model are optimized by solving (\ref{eq:min}) with the second loss term set to zero. Note that this makes the solution independent of the ordering of $\bb{k}$. 
  
  \item Run the TSP approximate solver to order the elements of the vector $\bb{k}$ and form the trajectory. This significantly reduces albeit still does not completely eliminate constraint violation, as can be seen in the input to stage 3 in Fig. \ref{fig:TSP-pipeline}. 
  
  \item The trajectory is fine-tuned to obey the constraints by further performing iteration on the loss in (\ref{eq:min}) this time with the constraints term. 
\end{enumerate}
It is important to note that unlike previous works using TSP for trajectory optimization, that \textit{modify} the target sampling density \citep{wang2012TSP,chauffert2014tsp}, in our work it is used only for the purpose of reordering the existing sampling points without any resampling along the trajectory (stage 3 above), therefore the sampling density remains unchanged after TSP.

\subsubsection{Multi-scale PILOT}
\label{multi-scale}
Experimentally, we observed that enforcing all the points as free optimization variables do not cause ``global" changes in the trajectory shape. One technique that proved effective to overcome this limitation is to learn only a subset of points that we refer to as \textit{control points}, with the rest of the trajectory interpolated using a differentiable cubic spline\footnote{A cubic spline is a spline constructed of piece-wise third-order polynomials which pass through a set of control points \citep{ferguson1964spline}.}. This method showed more ``global" changes but less ``randomly" spread points as proven to improve performance \citep{lustig2007sparse}. To compromise  between the ``global" and the ``local" effects, we adopted a multi-scale optimization approach. In this approach, we start from optimizing over a small set of control points per shot, and gradually increase their amount as the training progresses until all the trajectory points are optimized (e.g. starting from 60 points per shot with 3000 points and double their amount every 5 epochs).
It is important to note that we enforce the feasibility constraints on the trajectory \textit{after} the spline interpolation in order to keep the trajectory feasible during all the optimization steps.

\section{Experiments and Discussion}
\label{sec:experiments}

\subsection{Datasets}
\label{sec:datasets}
We used two datasets in the preparation of this article: the NYU \texttt{fastMRI} initiative database \citep{zbontar2018fastmri}, and the medical segmentation decathlon \citep{MedicalSegmentationDecathlon2019}. The \texttt{fastMRI} dataset contains raw knee MRI volumes with 2D spatial dimensions of $320\times320$. The \texttt{fastMRI} dataset consists of data obtained from multiple machines through two pulse sequences: proton-density and proton-density fat suppression. We included only the proton-density volumes in our dataset. Since our work focuses on designing $k$-space trajectories and the provided test set in \texttt{fastMRI} is already sub-sampled, we split the training set into two sets: one containing $484$ volumes ($17000$ slices) for training and validation (80/20 split), and $100$ volumes ($3500$ slices) for testing. Furthermore, the \texttt{fastMRI} dataset is acquired from multiple 3T scanners (Siemens Magnetom Skyra, Prisma, and Biograph mMR) as well as 1.5T machines (Siemens Magnetom Aera). All data were acquired through the fully-sampled Cartesian trajectories using the 2D TSE protocol. 

For the segmentation task, we used data obtained from the medical image segmentation \texttt{decathlon} challenge \citep{MedicalSegmentationDecathlon2019}. Within the \texttt{decathlon} challenge, we used the brain tumors dataset that contained $750$ 4D volumes (of $k$-space size $256\times256$) of multi-modal (FLAIR, T1w, T1gd \& T2w) brain MRI scans from patients diagnosed with either glioblastoma or lower-grade glioma. Gold standard annotations for all tumor regions in all scans were approved by expert board-certified neuroradiologists as detailed in \cite{MedicalSegmentationDecathlon2019}. Within this dataset, we used only the T2-weighted images. Since the goal of our experiment is to design trajectories that are optimal for segmenting tumors, we only considered images containing tumors, and used only the training set for all the experiments since they are the only ones that contain ground-truth segmentations. These data were split into two sets: one containing $400$ volumes ($26000$ slices) for training and validation, and the other one containing $84$ volumes ($5300$ slices) for testing. We emphasize that, due to the unavailability of the segmentation dataset that consists of true $k$-space data, for this experiment, we use the simulated $k$-space data obtained from the images. We believe that although the data are simulated, they can still be used in a proof-of-concept experiment.

\begin{table}[b]
    \centering
\caption{\small{\textbf{Comparison of fixed and learned trajectories in the single-shot setting for image reconstruction.} Presented are the PSNR and SSIM metrics of the fixed and learned trajectories for different decimation rates (\textrm{DR}). Reported baselines: fixed spiral trajectory (Spiral-Fixed), PILOT initialized with spiral (Spiral-PILOT), and the PILOT-TSP algorithm (PILOT-TSP).}}\label{tab-single-shot}
\begin{tabular}{|c|c|c|c|c|}
\hline
\textrm{DR}&Trajectory&Spiral-Fixed&Spiral-PILOT&PILOT-TSP\\
\hline
\multirow{2}{*}{10}&PSNR&28.16$\pm$1.43&31.87$\pm$1.47&\textbf{32.07$\pm$1.43}\\
&SSIM&0.734$\pm$0.041&0.814$\pm$0.035&\textbf{0.824$\pm$0.032}\\
\hline
\multirow{2}{*}{20}&PSNR&24.08$\pm$1.50&29.64$\pm$1.44&\textbf{30.25$\pm$1.48}\\
&SSIM&0.591$\pm$0.052&0.761$\pm$0.039&\textbf{0.776$\pm$0.040}\\
\hline
\multirow{2}{*}{80}&PSNR&21.05$\pm$1.61&24.31$\pm$1.49&\textbf{26.37$\pm$1.50}\\
&SSIM&0.478$\pm$0.049&0.598$\pm$0.048&\textbf{0.657$\pm$0.048}\\
\hline
\end{tabular}
\end{table} 

\subsection{Training settings}
We trained both the sub-sampling layer and the task model network with the Adam \citep{adam2014} solver. The learning rate was set to $0.001$ for the task model, while the sub-sampling layer was trained with learning rates varying between $0.001$ to $0.1$ in different experiments.
The parameters $\lambda_v$ and $\lambda_a$ were set to $0.1$.
We implemented the differentiable NUFFT and adjoint-NUFFT layers using \texttt{pytorch}'s \texttt{autograd} tools~\citep{NEURIPS2019_9015}, by adapting the code available in the \texttt{sigpy} package\footnote{https://github.com/mikgroup/sigpy}.  In both these layers, we use the Kaiser-Bessel kernel with oversampling factor set to 1.25, and interpolation kernel width (in terms of oversampled grid) set to 4.

\subsection{Physical constraints}
The following physical constraints were used in all our experiments: $G_\mathrm{max}=$ 40mT/m for the peak gradient, $S_\mathrm{max}=$ 200T/m/s for the maximum slew-rate, and $dt = 10\mu$sec for the sampling time.

\subsection{Image reconstruction}

\label{subsec:PILOTdiscuss}
In order to quantitatively evaluate our method, we use the peak signal-to-noise ratio (PSNR) and the structural-similarity measures (SSIM) \citep{ssim}, portraying both the pixel-to-pixel and perceptual similarity.
In all our experiments, we compare our algorithms to the baseline of training only the reconstruction model for measurements obtained with fixed handcrafted trajectories, henceforth referred to as ``fixed trajectory".

\subsubsection{Single-shot trajectories}
\vspace{-0.2cm}

\begin{wrapfigure}{r}{0.5\textwidth}
 \vspace{-0.4cm}
	\centering
	\includegraphics[width=0.5\textwidth]{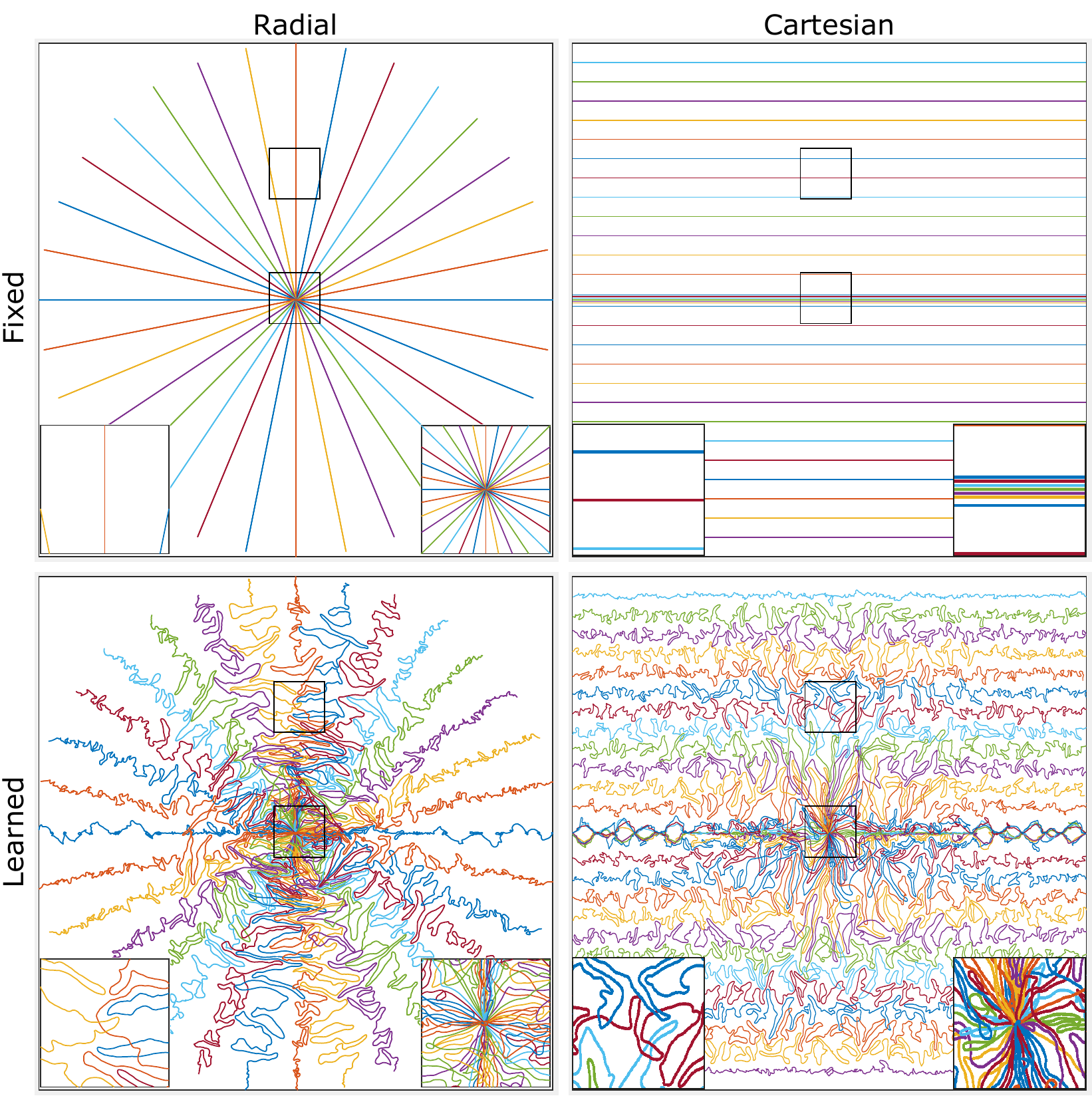} 
	\caption{\small{\textbf{Multi-shot PILOT trajectories.} Comparison of fixed (top) and learned (bottom) $k$-space trajectories for the radial (left) and Cartesian (right) initialization with $N_\textrm{shots}=16, 32$ respectively. The learned trajectories obey the following constraints: maximum gradient $G_\text{max} = 40 \, \text{mT/m} $, maximum slew-rate $S_\text{max} = 200 \,\text{T/m/s}$ and sampling interval $dt = 10 \, \mu s$.}}\label{fig:trajectories}
\end{wrapfigure}

For the single-shot setting, we used the spiral trajectory \citep{singleshotspiral2005, spiral1992} both as the baseline and the initialization to our model. We also used random initialization with PILOT-TSP, as described in Section \ref{sec:pilot-tsp}, initialized with .
In order to initialize feasible spirals that obey the hardware constraints, we used the variable-density spiral design toolbox available in \footnote{https://mrsrl.stanford.edu/~brian/vdspiral}, which produces spirals covering the $k$-space with the length of $m$ samples.
Table \ref{tab-single-shot} presents quantitative results comparing fixed spiral initialization to PILOT. The results demonstrate that PILOT outperforms the fixed trajectory across all decimation rates. As mentioned in Section \ref{subsec:subsampling}, the decimation rate (\textrm{DR}) is calculated as the ratio between the total grid size of the $k$-space ($n^2$) divided by the number of points sampled in the shot ($m$), i.e., $\textrm{DR} = n^2/m$.
The observed improvement is in the range of $3.26-5.56$ dB PSNR and $0.080-0.170$ SSIM points. 

Table \ref{tab-single-shot} also summarizes the quantitative results of PILOT-TSP. We observe that PILOT-TSP provides improvements in the range of $0.2-2.06$dB PSNR and $0.010-0. 059$ SSIM points when compared to the learned trajectory with the spiral initialization and a total of $3.91-6.17$dB PSNR and $0.090-0.185$ SSIM points overall improvement over fixed trajectories. Examples of PILOT-TSP trajectories can be seen in Figs \ref{fig:sparkling} and \ref{fig:segmentation}.

\subsubsection{Multi-shot trajectories}

\begin{table}
\vspace{-0.2cm}
	\centering
	
    \begin{tabular}{|c|c|c|}
    \hline
    &PSNR&SSIM\\
    \hline
    Fixed&29.09$\pm$1.43&0.741$\pm$0.040\\
    \hline
    PILOT (without multi-scale)&31.93$\pm$1.52&0.818$\pm$0.036\\
    \hline
    PILOT (with multi-scale)&\textbf{33.71$\pm$1.58}&\textbf{0.863$\pm$0.032}\\
    \hline
    \end{tabular}
    
	\caption{\small{\textbf{Multi-scale PILOT.} Comparison of PILOT with and without multi-scale optimization for the radial initialization with $N_\textrm{shots}=16$.}}\label{table:multi-scale}
\end{table}

In the multi-shot scenario, we used Cartesian, radial \citep{radial1973} and spiral \citep{spiral1992} trajectories both as the baseline and as the initialization for PILOT.
We denote the number of shots as $N_\textrm{shots}$, each containing $3000$ samples corresponding to the readout time of $30ms$.
As mentioned in Section \ref{multi-scale} we observed that using PILOT with multi-scale optimization lead to better performance. Therefore, in this work all multi-shot experiments use multi-scale optimization. Table \ref{table:multi-scale} shows a comparison between PILOT with and without multi-scale optimization.
Table \ref{tab-multi} presents the quantitative results comparing fixed trajectories to PILOT with the same initialization. The results demonstrate that in the multi-shot setting, as well, PILOT outperforms the fixed trajectory. In Fig. \ref{fig:trajectories} in the Cartesian setting, we can notice how the learned trajectory has much better coverage of the $k$-space which in turn led to huge improvement is in the range of $6.14-6.99$ dB PSNR and $0.161-0.172$ SSIM points. In the radial case, PILOT observes an improvement in the range of $3.31-4.99$ dB PSNR and $0.086-0.138$ SSIM points, and for spiral we see improvement in the range of $1.66-3.16$ dB PSNR and $0.029-0.084$ SSIM points.

More interestingly, for spiral initialization PILOT applied to the fewer shots setups ($4$ and $8$) surpasses the performance of the fixed trajectory setups with a higher number of shots ($8$ and $16$). For Cartesian and radial initializations, PILOT with as little as $8$ shots surpasses the $32$ fixed trajectory. This indicates that by using PILOT one can achieve about two to four times shorter acquisition without compromising the image quality (without taking into account additional acceleration potential due to the reconstruction network). \\ 

\begin{figure}
	\centering
	\includegraphics[width=0.6\textwidth]{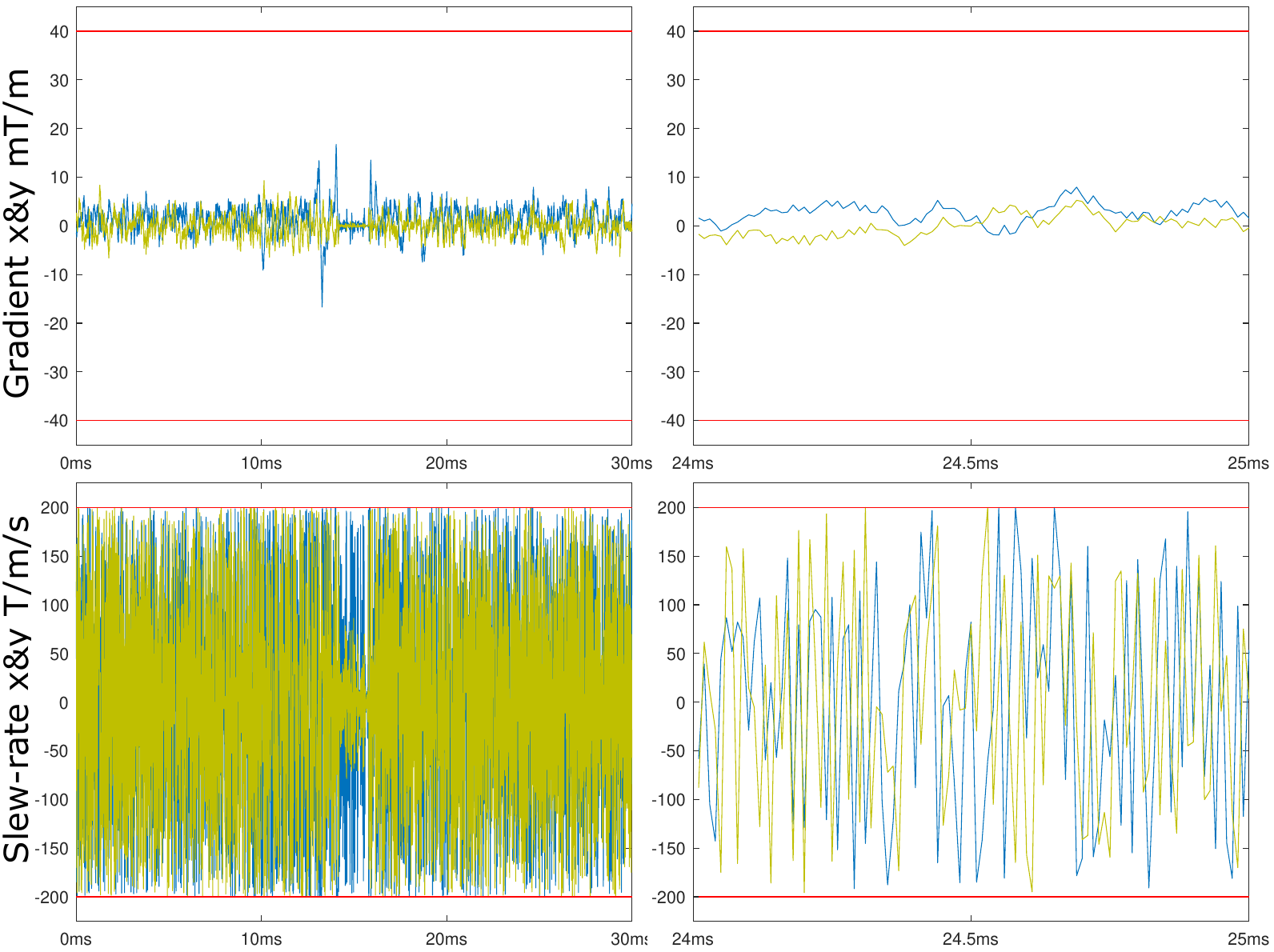} 
	\vspace{-0.2cm}
	\caption{\small Gradient (top) and slew-rate (bottom) plots of the learned $k$-space trajectory with radial initialization for $N_\textrm{shots}=16$. The right column depicts a zoom-in portion. The red lines are the constraints used in the optimization: maximum gradient $G_\text{max} = 40 \, \text{mT/m} $, maximum slew-rate $S_\text{max} = 200 \,\text{T/m/s}$.}\label{fig:grad+slew-rate}
\end{figure}

\begin{figure*}[htb!]
	\centering
	\includegraphics[width=1\textwidth]{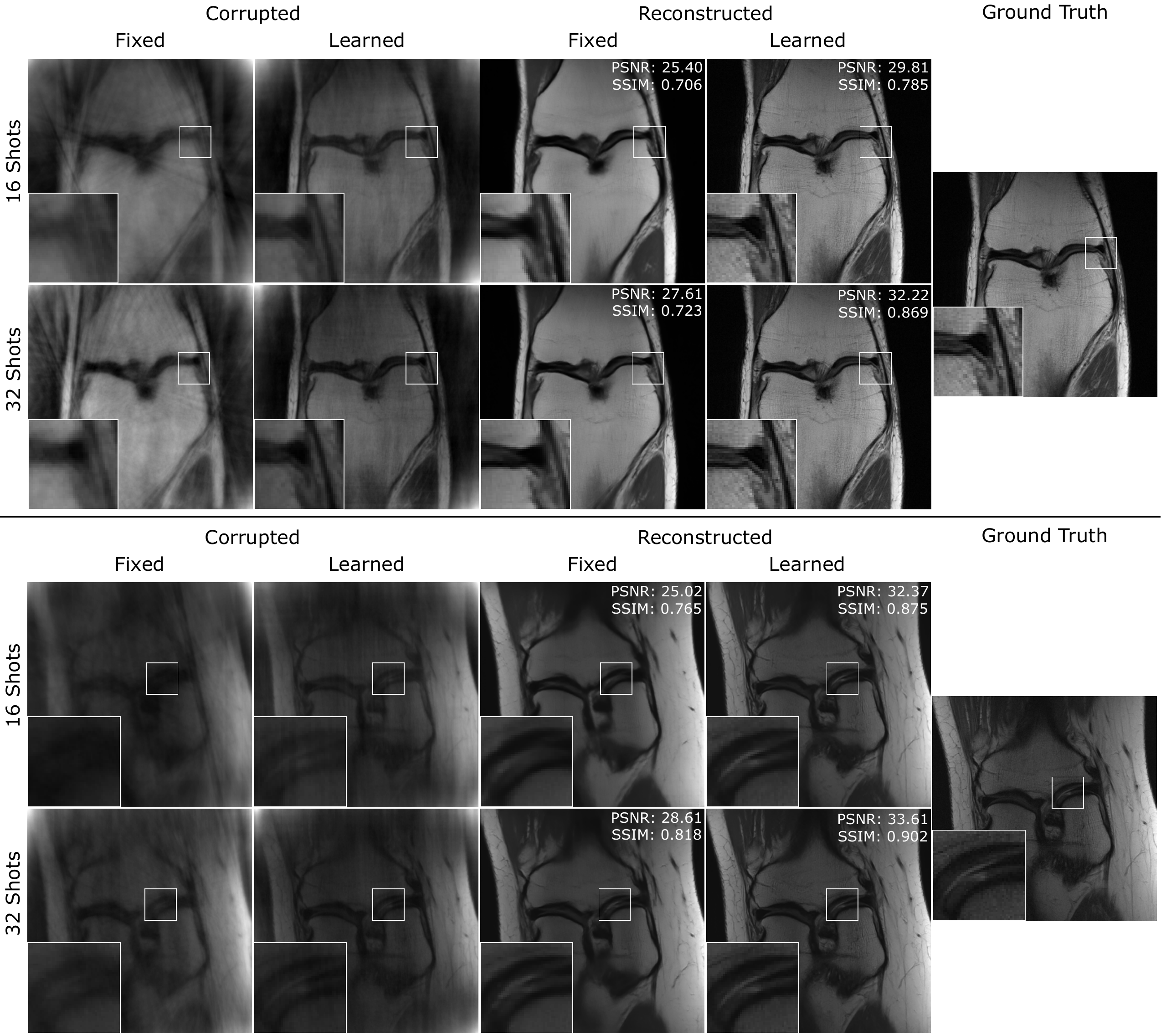}
	\caption{\small{\textbf{Image reconstruction with PILOT.} Compared are the fixed and learned trajectories for radial initialization with 16 (top row) and 32 shots (bottom row). Columns 1 \& 2 depict the images after resampling (root-sum-of-squares) and depicted in columns 3 \& 4 are the reconstructed images. The groundtruth image obtained from the fully sampled $k$-space is depicted in the rightmost column. PSNR and SSIM scores for the reconstructed images are reported alongside with zoom-in region displaying important anatomical details.}}\label{fig:vis-results}
\end{figure*}

An example of fixed and learned trajectories is presented in Figure \ref{fig:trajectories}. Gradient and slew rate of one of the learned trajectories can be seen in Fig. \ref{fig:grad+slew-rate}. Fig. \ref{fig:vis-results} shows a visual comparison of the distorted and reconstructed images using fixed and learned trajectories, it is interesting to observe the improvement in the distorted images which depend only on the trajectories.\\ 
The learned trajectories enjoy two main advantages over the traditional ("handcrafted") trajectories:
\begin{enumerate}
    \item The sampling points "adapt" (in a data-driven manner) to the sampling density of the specific organ, task, and task model (e.g. reconstruction method for reconstruction).
    \item The sampling points "jitter" around the trajectory in order to get better k-space coverage in multiple directions. This result resembles CAIPIRINHA-like zig-zag sampling that is traditionally handcrafted for parallel MR imaging \citep{breuer2005caipirinha}.
\end{enumerate}

 \begin{table*}
 \vspace{-0.3cm}
    \centering
\caption{\small{\textbf{Comparison of fixed and learned trajectories in the multi-shot setting for image reconstruction.} Presented are the PSNR and SSIM metrics of fixed and learned trajectories for different number of shots ($N_\textrm{shots} = 4, 8, 16, 32$). 
Reported baselines:
fixed spiral, radial, Cartesian trajectories (denoted by Spiral-Fixed, Radial-Fixed and Cartesian-Fixed, respectively);
PILOT initialized with spiral, radial and Cartesian trajectories (denoted by Spiral-PILOT, Radial-PILOT and Cartesian-PILOT, respectively).
}}\label{tab-multi}
\begin{tabular}{|c|c|c|c|c|}
\hline
$N_\textrm{shots}$ &\multicolumn{2}{|c|}{4}&
\multicolumn{2}{|c|}{8}\\
\hline
Trajectory&PSNR&SSIM&PSNR&SSIM\\
\hline
Cartesian-Fixed$^1$&-&-&24.45$\pm$1.44&0.622$\pm$0.044\\
Cartesian-PILOT$^1$&-&-&31.13$\pm$1.44&0.785$\pm$0.035\\
\hline
Radial-Fixed$^1$&-&-&27.24$\pm$1.41&0.686$\pm$0.044\\
Radial-PILOT$^1$&-&-&32.41$\pm$1.48&0.824$\pm$0.035\\
\hline
Spiral-Fixed&29.05$\pm$1.59&0.747$\pm$0.046&31.88$\pm$1.54&0.831$\pm$0.031\\
Spiral-PILOT&\textbf{32.21$\pm$ 1.50}&\textbf{0.831$\pm$0.036}&\textbf{34.54$\pm$1.46}&\textbf{0.876$\pm$0.033}\\
\hline \hline
$N_\textrm{shots}$&
\multicolumn{2}{|c|}{16}&
\multicolumn{2}{|c|}{32}\\
\hline
Trajectory&PSNR&SSIM&PSNR&SSIM\\
\hline
Cartesian-Fixed&24.93$\pm$1.48&0.639$\pm$0.043&27.78$\pm$1.44&0.711$\pm$0.042\\
Cartesian-PILOT&31.92$\pm$1.44&0.811$\pm$0.035&33.92$\pm$1.62&0.872$\pm$0.031\\

\hline
Radial-Fixed&29.09$\pm$1.43&0.741$\pm$0.040&31.09$\pm$1.48&0.798$\pm$0.037\\
Radial-PILOT&33.71$\pm$1.58&0.863$\pm$0.032&\textbf{34.40$\pm$1.58}&\textbf{0.884$\pm$0.029}\\
\hline
Spiral-Fixed$^2$&33.89$\pm$1.56&0.874$\pm$0.030&-&-\\
Spiral-PILOT$^2$&\textbf{35.55$\pm$1.59}&\textbf{0.903$\pm$0.026}&-&-\\
\hline
\end{tabular}
\begin{tablenotes}
\item [1] $^1$For $N_\textrm{shots}=4$, fixed Cartesian trajectories and fixed radial trajectories result in a very poor performance probably because of their poor coverage in the $k$-space, therefore we omit the results for these baselines for both fixed and learned trajectory baselines.
\item[2] $^2$Spiral trajectory with $N_\textrm{shots}=32$ resulted in less than 3000 sampling points per shot and was therefore omitted for a consistent comparison.
\end{tablenotes}
 \vspace{-0.3cm}
\end{table*}

\vspace{-0.2cm}

\subsection{Optimal trajectory design for segmentation}
We demonstrate the ability of the proposed methods to learn a trajectory optimized for a specific end-task on the tumor segmentation task. The details of the dataset used are described in Section \ref{sec:datasets}. 
We use the standard Dice similarity coefficient (DICE) and the intersection-over-union (IOU) metrics to quantify segmentation accuracy. 

\begin{figure*}[h!]
	\centering
	\includegraphics[width=1\textwidth]{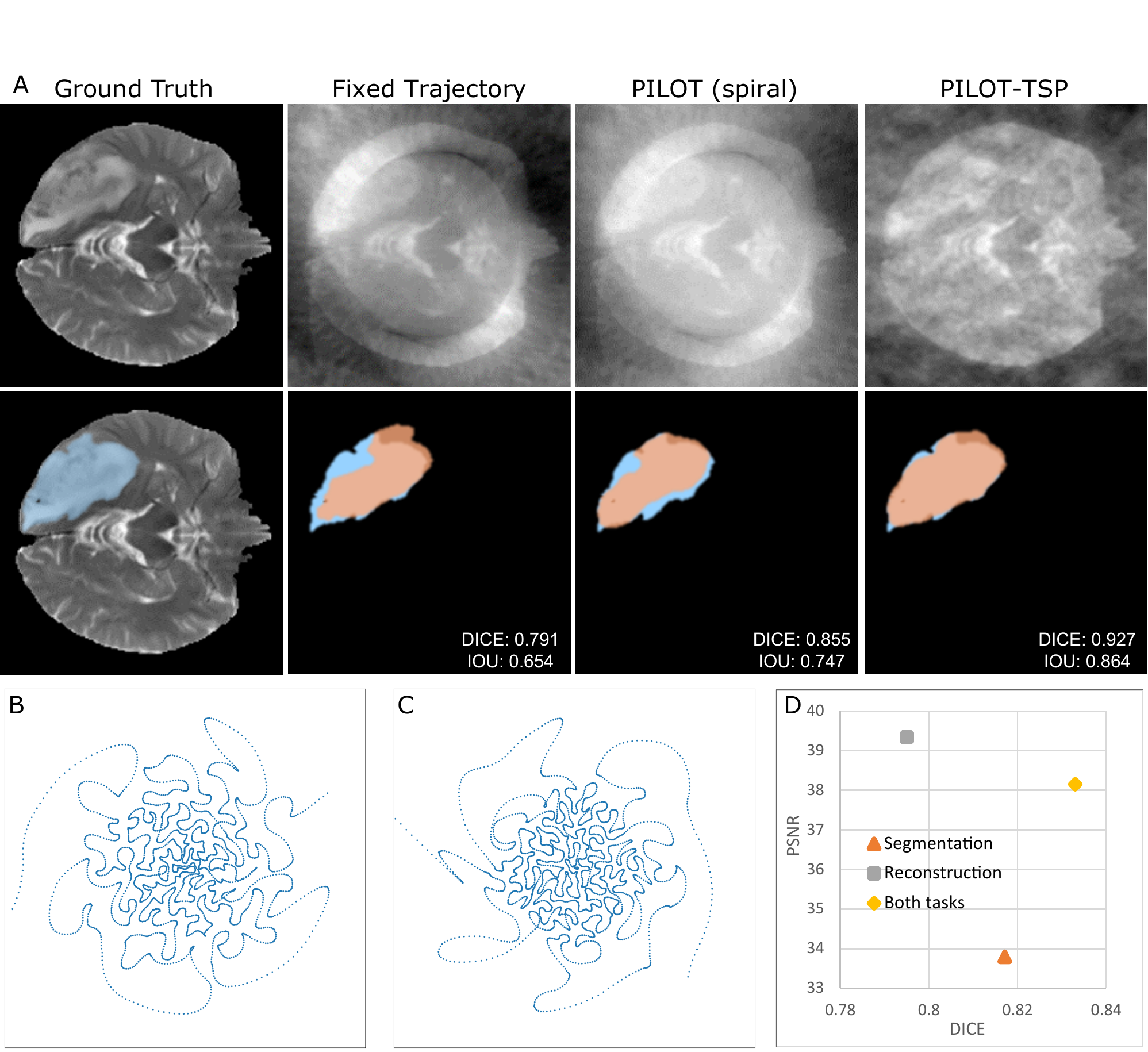} 
	\caption{\small{\textbf{Tumor segmentation.} (A) Compared are the fixed spiral trajectory (second column), PILOT with spiral initialization (third column), and PILOT-TSP (rightmost column). The first row depicts the MR images as obtained after the inverse Fourier transform with no further processing. The groundtruth image obtained from the fully sampled $k$-space is depicted in the leftmost column. The second row depicts the predicted segmentation map (red) and the groundtruth map (blue) with the corresponding DICE and IOU scores. (B, C) shows comparison of the learned trajectories optimized for different tasks: segmentation (B) and reconstruction (C). (D) Quantitative performance of trajectories optimized with different end-task (reconstruction, segmentation and both together) on both tasks - reconstruction (PSNR) and segmentation (DICE)}} \label{fig:segmentation}
\end{figure*}

The results for segmentation are reported in Table \ref{tab-seg}. We observe that the learned trajectories achieve an improvement of 
$0.015-0.044$ DICE points and $0.018-0.056$ IOU points when compared to the fixed trajectories they were initialized with. Visual comparison of the described results is available in Fig.\ref{fig:segmentation}.  
We emphasize that since the dataset for the segmentation task was likely post-processed and therefore the retrospective data acquisition simulation is not as accurate as in the case of reconstruction task where the raw multi-channel $k$-space data was available. Nevertheless, the margins achieved in reconstruction experiments encourage us to believe that similar margins would be obtained on the segmentation task as well when trained with raw $k$-space data as the input. 

\begin{table}
    \centering
\addtolength{\tabcolsep}{1pt}
\caption{\small{\textbf{Comparison of fixed and learned trajectories in the single-shot setting for the tumor segmentation task.} Presented are the DICE and IOU metrics of the fixed and learned trajectories for different decimation rates (\textrm{DR}). Reported baselines: fixed spiral trajectory (Spiral-Fixed), PILOT initialized with spiral (Spiral-PILOT), and the PILOT-TSP algorithm (PILOT-TSP).
Three different decimation rates are reported.}}\label{tab-seg}
\begin{tabular}{|c|c|c|c|c|}
\hline
\textrm{DR}&Trajectory&Spiral-Fixed&Spiral-PILOT&PILOT-TSP\\
\hline
\multirow{2}{*}{10}&DICE&0.810$\pm$0.012&\textbf{0.825$\pm$0.011}&0.816$\pm$0.009\\
&IOU&0.719$\pm$0.015&\textbf{0.737$\pm$0.014}&0.726$\pm$0.012\\
\hline
\multirow{2}{*}{20}&DICE&0.774$\pm$0.016&0.811$\pm$0.016&\textbf{0.818$\pm$0.017}\\
&IOU&0.669$\pm$0.017&0.716$\pm$0.016&\textbf{0.725$\pm$0.016}\\
\hline
\multirow{2}{*}{80}&DICE&0.720$\pm$0.022&0.735$\pm$0.019&\textbf{0.754$\pm$0.018}\\
&IOU&0.611$\pm$0.019&0.626$\pm$0.016&\textbf{0.639$\pm$0.017}\\
\hline
\end{tabular}
\end{table}

\subsection{Cross-task and multi-task learning}

In order to verify end-task optimality of the found solution, we evaluated the learned models on tasks they were not learned on. Specifically, we trained PILOT-TSP with the same dataset once on the segmentation task and another time on the reconstruction task. We then swapped the obtained trajectories between the tasks, fixed them, and re-trained only the task model (U-Net) for each task (while keeping the trajectory as it was learned for the other task). The results for this experiment are reported in Fig. \ref{fig:segmentation}. We observe that the trajectories obtained from segmentation are optimal for that task alone and perform worse on reconstruction, and vice-versa. From the output trajectories learned for segmentation and reconstruction, visualized in Fig \ref{fig:segmentation}, we can see that the reconstruction trajectory is more centered around the DC frequency, whereas the segmentation trajectory tries to cover higher frequencies. This could be explained by the fact that most of the image information is contained within the lower frequencies in the $k$-space, therefore covering them, contributes more to the pixel-wise similarity measure of the reconstruction task. For the segmentation task, contrarily, the edges and the structural information present in the higher frequencies, is more critical for the success of the task.  
This is an interesting observation because it paves the way to design certain accelerated MRI protocols that are optimal for a given end-task that is not necessarily image reconstruction. 

Another important use can be to jointly learn several tasks such as segmentation and reconstruction in order to find the best segmentation while preserving the ability to reconstruct a meaningful human-intelligible image. For this purpose, we build a model with one encoder and two separate decoders (separated at the bottleneck layer of the U-Net), each for a different task. The loss function is defined to be the sum of both the reconstruction loss (from the first decoder) and the segmentation loss (pixel-wise cross-entropy, from the second decoder). The results for this experiment are reported in Fig. \ref{fig:segmentation}. Interestingly, we can observe that the presence of the reconstruction tasks aids segmentation as well. This is also consistent with the results observed in \cite{sun2018reconsegment}.

\subsection{Comparison with prior art}

\begin{wrapfigure}{r}{0.5\textwidth}
	\centering
	\vspace{-0.6cm}
	\includegraphics[width=0.5\columnwidth]{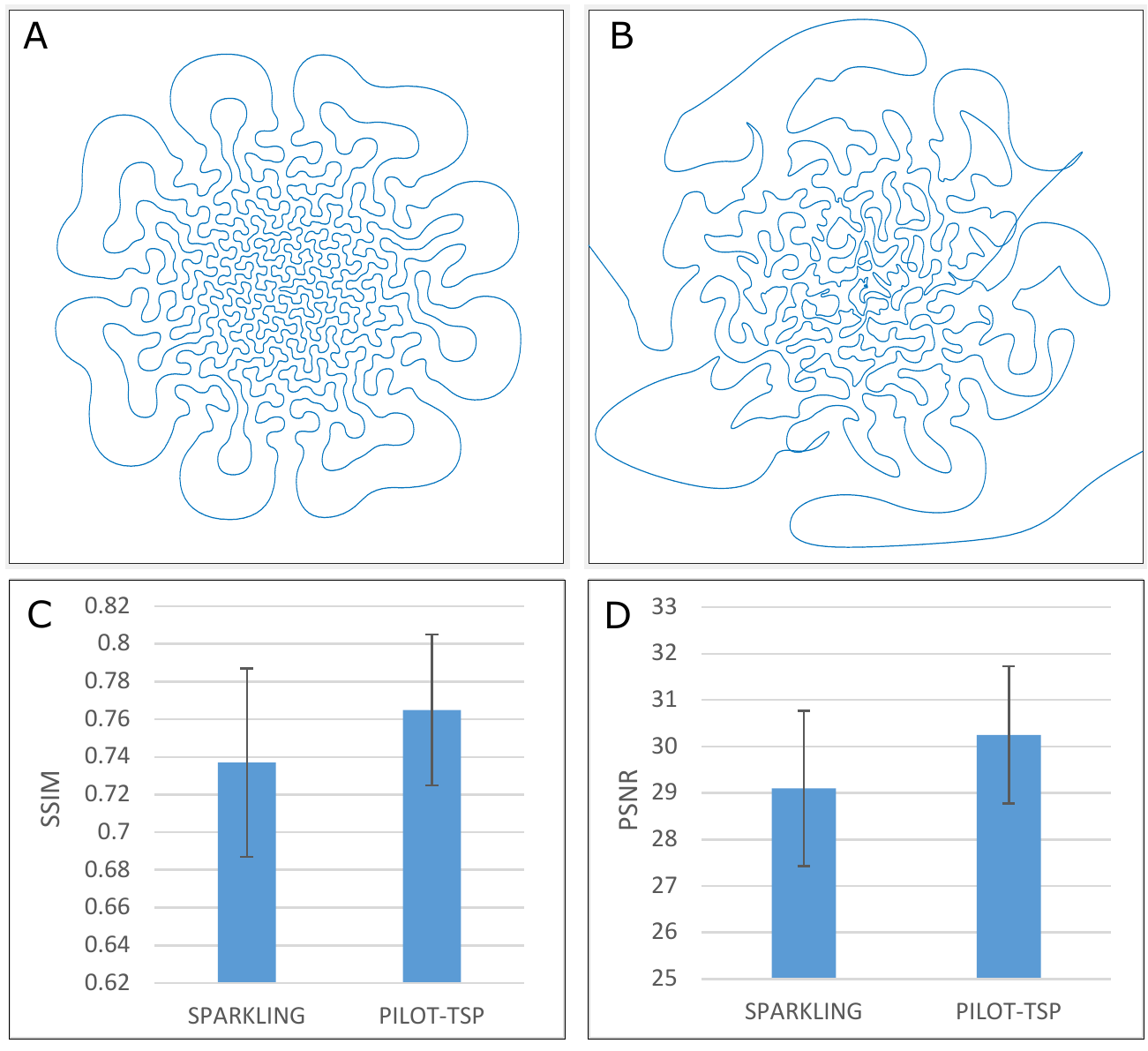} 
	\caption{\small{Comparison of SPARKLING (A) and PILOT-TSP (B) trajectories for decimation rate of 20 with evaluation of the reconstruction using U-Net (C, D).}}\label{fig:sparkling}
	\vspace{-0.4cm}
\end{wrapfigure}

SPARKLING \citep{sparkling2019Lazarus} is the state-of-the-art method for single-shot 2D k-space optimization under gradient amplitude and slew constraints, we use it for comparative evaluation of PILOT. A single-shot trajectory corresponding to decimation rate $20$ was generated using the variable density sampling density suggested in the paper, using the implementation provided in \cite{optimaltransport2018leo}, an extension of SPARKLING\footnote{Due to implementation limitations of the released SPARKLING code, we were able to produce only single-shot trajectories, and therefore the comparison of SPARKLING to PILOT is limited to this scenario. All the hyper-parameters were set to their default values.}. We tested the performance of both trajectories with U-Net as the reconstruction methods, trained for each trajectory. Figure \ref{fig:sparkling} displays SPARKLING and PILOT-TSP generated trajectories and the quantitative results. As can be inferred from the results, PILOT outperforms SPARKLING with $1.15$ dB PSNR and $0.030$ SSIM points. We assume the reason for the superior performance of PILOT when compared to SPARKLING is mainly due to PILOT's data-driven approach for selecting the sampling points. SPARKLING tries to fit the trajectory to a ``handcrafted" \textit{pre-determined} sampling density, where as in PILOT the sampling density is implicitly optimized together with the trajectory to maximize the performance of the end task.

\subsection{Resilience to noise}
\begin{figure*}[t]
	\centering
	\includegraphics[width=\textwidth]{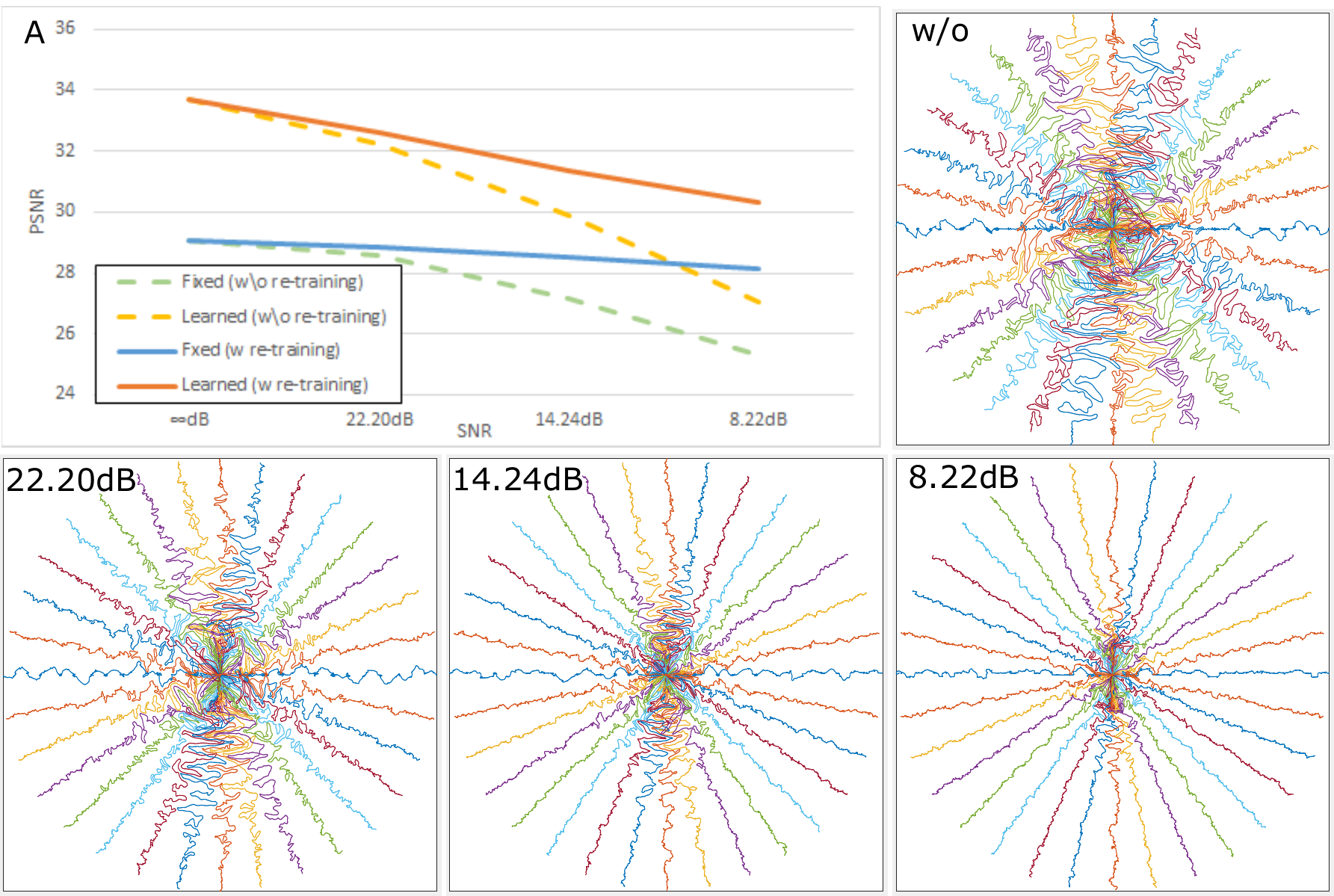} 
	\caption{\small{Comparison of PILOT with radial initialization with and without accounting for lower SNR. The dashed lines (yellow and green) in sub-figure A correspond to the performance when the SNR decay was emulated only during inference, i.e., the models were trained without accounting for the lower SNR. The continuous lines (orange and blue) in sub-figure A presents the performance of the models when \textit{trained} with lower SNR. The effect of the SNR on the learned trajectories can be seen in the trajectories plots, the upper-right trajectory trained without noise and the bottom trajectories were trained with different levels of SNR.}}\label{fig:SNR}
\end{figure*}

In order to test the effect of measurement noise on the performance of PILOT trajectories, we designed the following twofold experiment: (1) emulated the lower SNR only in the test samples using trajectories trained without the noise (\textit{w/o retraining}); and (2) emulating the lower SNR both in train and test samples (\textit{with retraining}).\\
\textbf{SNR emulation}. In order to emulate measurement noise we added white Gaussian noise to the real and imaginary parts of each of the input channels. We used noise with higher variance to emulate lower SNR. The results are summarized in Fig. \ref{fig:SNR}. We observe the relative advantage of PILOT in both scenarios, with and without training over low SNR samples, while for the latter the learned trajectory and model are more resilient in the presence of noise. We assume that this improvement is both due to the adaptation of the reconstruction network and due to trajectory adaptation to the lower SNR. Learned trajectories in the presence of different SNR values can be seen in Fig. \ref{fig:SNR}, we can see that in the higher SNR case the sampling points are more dispersed  over the $k$-space than in the lower SNR case. This is probably due to the need to average many close sampling points in order to counter the noise.

\section{Conclusion and future directions}
\label{sec:Conclusion}

We have demonstrated, as a proof-of-concept, that learning the $k$-space sampling trajectory simultaneously with an image reconstruction neural model improves the end image quality of an MR imaging system. To the best of our knowledge, the proposed PILOT algorithm is the first to do end-to-end learning over the space of all physically feasible MRI 2D $k$-space trajectories. We also showed that end-to-end learning is possible with other tasks such as segmentation.
On both tasks and across decimation rates and initializations PILOT manages to learn physically feasible trajectories which demonstrates significant improvement over the fixed counterparts.
We proposed two extensions to the PILOT framework, namely, multi-scale PILOT and PILOT-TSP in order to partially alleviate the local minima that are prevalent in trajectory optimization. 
We further studied the robustness of our learned trajectories and reconstruction model in the presence of SNR decay.\\
Below we list some limitations of the current work and some future directions:
\begin{itemize}
  \item In this paper, we only validated our method through retrospective acquisitions, we hope to validate it prospectively by deploying the learned trajectories on a real MRI machine.
  \item We were unable to obtain a segmentation dataset containing the raw $k$-space data, we plan to obtain such datasets and evaluate the performance of our methods on them.
  \item Adding echo time (TE) constraints in order to control the received contrast.
  \item Considering a more accurate physical model accounting for eddy current effects (e.g. gradient impulse response function (GIRF) model \citep{vannesjo2013GIRF}).
  \item We did not evaluate the robustness of our trajectories to off-resonance effects. Preliminary simulation results in \cite{sparkling2019Lazarus} suggested that similarly optimized trajectories are more robust to off-resonance effects when compared to spiral. In future work, we will evaluate the robustness of PILOT by performing similar simulations.
  \item Extending PILOT-TSP to multi-shot trajectories. 
\end{itemize}


\acks{This work was supported by the ERC CoG EARS.}

%
\ethics{The work follows appropriate ethical standards in conducting research and writing the manuscript, following all applicable laws and regulations regarding treatment of animals or human subjects.}

\coi{We declare we don't have conflicts of interest.}


\vskip 0.2in
\bibliography{refs}

\appendix 

\section*{Appendix A.}
\label{appendix-A}
In what follows we describe the implementation of the differentiable interpolation step used in the sub-sampling and regridding layers.
Let us assume we know the value of a function $f$ on a discrete grid, $X= \{i\in1...n\} , f(i)=y_i$ and we would like to evaluate the value of $f$ at a point $z$ that does not lie on the grid. We perform this operation in two steps:
\begin{enumerate}
    \item Find the points in $X$ that satisfy $\{j, j+1 \in X \, |  \, j = \text{floor}(z)\}$. This step is indeed piecewise differentiable with respect to $z$: the gradient is defined and equals zero outside the set $X$ and is not defined otherwise. Therefore the gradient is zero almost everywhere.
    \item Now we define a function $f_{cont}$, an interpolation function, to estimate the value of the function $f$ at $z$, $f_{cont}(z)=(z-j)f(j) + (j+1-z)f(j+1)$. This step is differentiable with respect to z.
\end{enumerate}
The function $f_{cont}$ is the function defined on the real domain and it is equal to the discrete function $f$ when the input is an integer. Let us now evaluate the derivative of $f_{cont}$ w.r.to $z$
$$
\frac{\partial f_{cont}}{\partial z} = f(j) - f(j+1) + (z-j)f'(j)\frac{\partial j}{\partial z} + (j+1-z)f'(j+1)\frac{\partial j}{\partial z}
$$
and since, $j = \text{floor}(z)$ has $\frac{\partial j}{\partial z}=0,  \forall z \notin X$ and undefined elsewhere we assume $\frac{\partial j}{\partial z}=0$. This yields
$\frac{\partial f_{cont}}{\partial z}=f(j) - f(j+1)$.
While in this example we considered the simple case of linear interpolation, the general logic applies for higher-order or kernel-based interpolation techniques as well. In our implementation, however, we did not use this kind of direct approach for the calculation of the gradient, instead we used the \texttt{autograd} engine available in \texttt{pytorch}~\citep{NEURIPS2019_9015} for this purpose.

\end{document}